% Routine to guarantee that this file is input only once
\catcode`\@=11
\expandafter\ifx\csname @iasmacros\endcsname\relax
	\global\let\@iasmacros=\par
\else	\immediate\write16{}
	\immediate\write16{Warning:}
	\immediate\write16{You have tried to input iasmacros more than once.}
	\immediate\write16{}
	\endinput
\fi
\catcode`\@=12

% Set up font size commands and \baselinestretch command
%\input iasfonts

% Some alternative font names
\def\rmb{\seventeenrm}

% Simple spacing commands
\def\singlespace{\baselineskip=\normalbaselineskip}
\def\halfspace{\baselineskip=1.5\normalbaselineskip}
\def\doublespace{\baselineskip=2\normalbaselineskip}

% Macros for references and abstracts

\def\AB{\bigskip\parindent=40pt
        \centerline{\bf ABSTRACT}\medskip\halfspace\narrower}
\def\AE{\bigskip\nonarrower\doublespace}
\def\nonarrower{\advance\leftskip by-\parindent
	\advance\rightskip by-\parindent}

% Useful commands

\def\boxit#1{\vbox{\hrule\hbox{\vrule\kern3pt
	\vbox{\kern3pt#1\kern3pt}\kern3pt\vrule}\hrule}}

% Special symbols
\def\hence{\leavevmode\hbox{\bf .\raise5.5pt\hbox{.}.} }

\def\dalemb#1#2{{\vbox{\hrule height.#2pt
	\hbox{\vrule width.#2pt height#1pt \kern#1pt \vrule width.#2pt}
	\hrule height.#2pt}}}
\def\gtorder{\mathrel{\raise.3ex\hbox{$>$}\mkern-14mu
             \lower0.6ex\hbox{$\sim$}}}
\def\ltorder{\mathrel{\raise.3ex\hbox{$<$}\mkern-14mu
             \lower0.6ex\hbox{$\sim$}}}

% For twoup output
\newdimen\fullhsize
\newbox\leftcolumn
\def\twoup{\hoffset=-.5in \voffset=-.25in
  \hsize=4.75in \fullhsize=10in \vsize=6.9in
  \def\fullline{\hbox to\fullhsize}
  \let\lr=L
  \output={\if L\lr
        \global\setbox\leftcolumn=\columnbox\global\let\lr=R \advancepageno
      \else \doubleformat \global\let\lr=L\fi
    \ifnum\outputpenalty>-20000 \else\dosupereject\fi}
  \def\doubleformat{\shipout\vbox{
    \fullline{\box\leftcolumn\hfil\columnbox}\advancepageno}}
  \def\columnbox{\leftline{\vbox{\makeheadline\pagebody\makefootline}}}
  \tolerance=1000 }
\catcode`\@=11					% To make protected \def's

%************************************************************
%*
%*		Font set-up
%*
%************************************************************

%************** 5-point fonts *******************************

\font\fiverm=cmr5				% roman
\font\fivemi=cmmi5				% math italic
\font\fivesy=cmsy5				% math symbols
\font\fivebf=cmbx5				% bold face

\skewchar\fivemi='177
\skewchar\fivesy='60

%************** 6-point fonts *******************************

\font\sixrm=cmr6				% roman
\font\sixi=cmmi6				% math italic
\font\sixsy=cmsy6				% math symbols
\font\sixbf=cmbx6				% bold face

\skewchar\sixi='177
\skewchar\sixsy='60

%************** 7-point fonts *******************************

\font\sevenrm=cmr7				% roman
\font\seveni=cmmi7				% math italic
\font\sevensy=cmsy7				% math symbols
\font\sevenit=cmti7				% italic
\font\sevenbf=cmbx7				% bold face

\skewchar\seveni='177
\skewchar\sevensy='60

%************** 8-point fonts *******************************

\font\eightrm=cmr8				% roman
\font\eighti=cmmi8				% math italic
\font\eightsy=cmsy8				% math symbols
\font\eightit=cmti8				% italic
				% slanted
\font\eightbf=cmbx8				% bold face
				% typewriter
				% sans serif

\skewchar\eighti='177
\skewchar\eightsy='60

%************** 9-point fonts *******************************

\font\ninei=cmmi9
\font\ninesy=cmsy9

\skewchar\ninei='177
\skewchar\ninesy='60

%************** 10-point fonts ******************************

\font\tenrm=cmr10				% roman
\font\teni=cmmi10				% math italic
\font\tensy=cmsy10				% math symbols
\font\tenex=cmex10				% math extension
\font\tenit=cmti10				% italic
\font\tensl=cmsl10				% slanted
\font\tenbf=cmbx10				% bold face
\font\tentt=cmtt10				% typewriter
\font\tenss=cmss10				% sans serif
\font\tensc=cmcsc10				% small caps
\font\tenbi=cmmib10				% bold math

\skewchar\teni='177
\skewchar\tenbi='177
\skewchar\tensy='60

\def\tenpoint{\ifmmode\err@badsizechange\else
	\textfont0=\tenrm \scriptfont0=\sevenrm \scriptscriptfont0=\fiverm
	\textfont1=\teni  \scriptfont1=\seveni  \scriptscriptfont1=\fivemi
	\textfont2=\tensy \scriptfont2=\sevensy \scriptscriptfont2=\fivesy
	\textfont3=\tenex \scriptfont3=\tenex   \scriptscriptfont3=\tenex
	\textfont4=\tenit \scriptfont4=\sevenit \scriptscriptfont4=\sevenit
	\textfont5=\tensl
	\textfont6=\tenbf \scriptfont6=\sevenbf \scriptscriptfont6=\fivebf
	\textfont7=\tentt
	\textfont8=\tenbi \scriptfont8=\seveni  \scriptscriptfont8=\fivemi
	\def\rm{\tenrm\fam=0 }%
	\def\it{\tenit\fam=4 }%
	\def\sl{\tensl\fam=5 }%
	\def\bf{\tenbf\fam=6 }%
	\def\tt{\tentt\fam=7 }%
	\def\ss{\tenss}%
	\def\sc{\tensc}%
	\def\bmit{\fam=8 }%
	\rm\setparameters\setbaselines\fi}

%************** 12-point fonts ******************************

\font\twelverm=cmr12				% roman
\font\twelvei=cmmi12				% math italic
\font\twelvesy=cmsy10	scaled\magstep1		% math symbols
\font\twelveex=cmex10	scaled\magstep1		% math extension
\font\twelveit=cmti12				% italic
\font\twelvesl=cmsl12				% slanted
\font\twelvebf=cmbx12				% bold face
\font\twelvett=cmtt12				% typewriter
\font\twelvess=cmss12				% sans serif
\font\twelvesc=cmcsc10	scaled\magstep1		% small caps
\font\twelvebi=cmmib10	scaled\magstep1		% bold math

\skewchar\twelvei='177
\skewchar\twelvebi='177
\skewchar\twelvesy='60

\def\twelvepoint{\ifmmode\err@badsizechange\else
	\textfont0=\twelverm \scriptfont0=\eightrm \scriptscriptfont0=\sixrm
	\textfont1=\twelvei  \scriptfont1=\eighti  \scriptscriptfont1=\sixi
	\textfont2=\twelvesy \scriptfont2=\eightsy \scriptscriptfont2=\sixsy
	\textfont3=\twelveex \scriptfont3=\tenex   \scriptscriptfont3=\tenex
	\textfont4=\twelveit \scriptfont4=\eightit \scriptscriptfont4=\sevenit
	\textfont5=\twelvesl
	\textfont6=\twelvebf \scriptfont6=\eightbf \scriptscriptfont6=\sixbf
	\textfont7=\twelvett
	\textfont8=\twelvebi \scriptfont8=\eighti  \scriptscriptfont8=\sixi
	\def\rm{\twelverm\fam=0 }%
	\def\it{\twelveit\fam=4 }%
	\def\sl{\twelvesl\fam=5 }%
	\def\bf{\twelvebf\fam=6 }%
	\def\tt{\twelvett\fam=7 }%
	\def\ss{\twelvess}%
	\def\sc{\twelvesc}%
	\def\bmit{\fam=8 }%
	\rm\setparameters\setbaselines\fi}

%************** 14-point fonts ******************************

\font\fourteenrm=cmr12	scaled\magstep1		% roman
\font\fourteeni=cmmi12	scaled\magstep1		% math italic
\font\fourteensy=cmsy10	scaled\magstep2		% math symbols
\font\fourteenex=cmex10	scaled\magstep2		% math extension
\font\fourteenit=cmti12	scaled\magstep1		% italic
\font\fourteensl=cmsl12	scaled\magstep1		% slanted
\font\fourteenbf=cmbx12	scaled\magstep1		% bold face
\font\fourteentt=cmtt12	scaled\magstep1		% typewriter
\font\fourteenss=cmss12	scaled\magstep1		% sans serif
\font\fourteensc=cmcsc10 scaled\magstep2	% small caps
\font\fourteenbi=cmmib10 scaled\magstep2	% bold math

\skewchar\fourteeni='177
\skewchar\fourteenbi='177
\skewchar\fourteensy='60

\def\fourteenpoint{\ifmmode\err@badsizechange\else
	\textfont0=\fourteenrm \scriptfont0=\tenrm \scriptscriptfont0=\sevenrm
	\textfont1=\fourteeni  \scriptfont1=\teni  \scriptscriptfont1=\seveni
	\textfont2=\fourteensy \scriptfont2=\tensy \scriptscriptfont2=\sevensy
	\textfont3=\fourteenex \scriptfont3=\tenex \scriptscriptfont3=\tenex
	\textfont4=\fourteenit \scriptfont4=\tenit \scriptscriptfont4=\sevenit
	\textfont5=\fourteensl
	\textfont6=\fourteenbf \scriptfont6=\tenbf \scriptscriptfont6=\sevenbf
	\textfont7=\fourteentt
	\textfont8=\fourteenbi \scriptfont8=\tenbi \scriptscriptfont8=\seveni
	\def\rm{\fourteenrm\fam=0 }%
	\def\it{\fourteenit\fam=4 }%
	\def\sl{\fourteensl\fam=5 }%
	\def\bf{\fourteenbf\fam=6 }%
	\def\tt{\fourteentt\fam=7}%
	\def\ss{\fourteenss}%
	\def\sc{\fourteensc}%
	\def\bmit{\fam=8 }%
	\rm\setparameters\setbaselines\fi}

%************** Miscellaneous big fonts *********************

\font\seventeenrm=cmr10 scaled\magstep3		% roman
		% bold face

%************************************************************
%*
%*		Parameter initialization
%*
%************************************************************

\newdimen\rp@
\newcount\@basestretchnum
\newskip\@baseskip
\newskip\headskip
\newskip\footskip

% Routine to set page parameters

\def\setparameters{\rp@=.1em
	\headskip=24\rp@
	\footskip=\headskip
	\delimitershortfall=5\rp@
	\nulldelimiterspace=1.2\rp@
	\scriptspace=0.5\rp@
	\abovedisplayskip=10\rp@ plus3\rp@ minus5\rp@
	\belowdisplayskip=10\rp@ plus3\rp@ minus5\rp@
	\abovedisplayshortskip=5\rp@ plus2\rp@ minus4\rp@
	\belowdisplayshortskip=10\rp@ plus3\rp@ minus5\rp@
	\normallineskip=\rp@
	\lineskip=\normallineskip
	\normallineskiplimit=0pt
	\lineskiplimit=\normallineskiplimit
	\jot=3\rp@
	\setbox0=\hbox{\the\textfont3 B}\p@renwd=\wd0
	\skip\footins=12\rp@ plus3\rp@ minus3\rp@
	\skip\topins=0pt plus0pt minus0pt}

% Special routine to scale \baselineskip

\def\setbaselines{\maxdepth=4\rp@\baselinestretch=\@basestretchnum}

% The \baselinestretch command

\def\baselinestretch{\afterassignment\@basestretch\@basestretchnum}
\def\@basestretch{%
	\@baseskip=12\rp@ \divide\@baseskip by1000
	\normalbaselineskip=\@basestretchnum\@baseskip
	\baselineskip=\normalbaselineskip
	\bigskipamount=\the\baselineskip
		plus.25\baselineskip minus.25\baselineskip
	\medskipamount=.5\baselineskip
		plus.125\baselineskip minus.125\baselineskip
	\smallskipamount=.25\baselineskip
		plus.0625\baselineskip minus.0625\baselineskip
	\setbox\strutbox=\hbox{\vrule height.708\baselineskip
		depth.292\baselineskip width0pt }}

%************************************************************
%*
%*		Modifications to PLAIN.TEX
%*
%************************************************************

% Modifications to PLAIN routines to handle scaling of page parameters

\def\makeheadline{\vbox to0pt{\baselinestretch=1000
	\vskip-\headskip \vskip1.5pt
	\line{\vbox to\ht\strutbox{}\the\headline}\vss}\nointerlineskip}

\def\makefootline{\baselineskip=\footskip\line{\the\footline}}

\def\big#1{{\hbox{$\left#1\vbox to8.5\rp@ {}\right.\n@space$}}}
\def\Big#1{{\hbox{$\left#1\vbox to11.5\rp@ {}\right.\n@space$}}}
\def\bigg#1{{\hbox{$\left#1\vbox to14.5\rp@ {}\right.\n@space$}}}
\def\Bigg#1{{\hbox{$\left#1\vbox to17.5\rp@ {}\right.\n@space$}}}

% Modifications to PLAIN to handle bold math

\mathchardef\alpha="710B
\mathchardef\beta="710C
\mathchardef\gamma="710D
\mathchardef\delta="710E
\mathchardef\epsilon="710F
\mathchardef\zeta="7110
\mathchardef\eta="7111
\mathchardef\theta="7112
\mathchardef\iota="7113
\mathchardef\kappa="7114
\mathchardef\lambda="7115
\mathchardef\mu="7116
\mathchardef\nu="7117
\mathchardef\xi="7118
\mathchardef\pi="7119
\mathchardef\rho="711A
\mathchardef\sigma="711B
\mathchardef\tau="711C
\mathchardef\upsilon="711D
\mathchardef\phi="711E
\mathchardef\chi="711F
\mathchardef\psi="7120
\mathchardef\omega="7121
\mathchardef\varepsilon="7122
\mathchardef\vartheta="7123
\mathchardef\varpi="7124
\mathchardef\varrho="7125
\mathchardef\varsigma="7126
\mathchardef\varphi="7127
\mathchardef\imath="717B
\mathchardef\jmath="717C
\mathchardef\ell="7160
\mathchardef\wp="717D
\mathchardef\partial="7140
\mathchardef\flat="715B
\mathchardef\natural="715C
\mathchardef\sharp="715D

%************************************************************
%*
%*		Initialization
%*
%************************************************************

\def\err@badsizechange{%
	\immediate\write16{--> Size change not allowed in math mode, ignored}}

\baselinestretch=1000
\tenpoint

\catcode`\@=12					% Restore @ sign
\twelvepoint
\doublespace
{\nopagenumbers{
\rightline{IASSNS-HEP-97/51; DAMTP-97-99}
\rightline{~~~September, 1997}
\bigskip\bigskip
\centerline{\rmb Corrections to the Emergent Canonical Commutation Relations} 
\centerline{\rmb Arising in the Statistical Mechanics of Matrix Models}
\medskip
\centerline{\it Stephen L. Adler}
\centerline{\bf Institute for Advanced Study}
\centerline{\bf Princeton, NJ 08540}
\medskip
\centerline{\it Achim Kempf}
\centerline{\bf Department of Applied Mathematics and Theoretical Physics}
\centerline{\bf University of Cambridge}
\centerline{\bf Cambridge CB3 9EW, U.K.}
\medskip

%\leftline{{\it Short title:} short title}
\bigskip\bigskip
\leftline{\it Send correspondence to:}
\medskip
{\singlespace\leftline{Stephen L. Adler}
\leftline{Institute for Advanced Study}
\leftline{Olden Lane, Princeton, NJ 08540}
\leftline{Phone 609-734-8051; FAX 609-924-8399; email adler@ias.edu}}
\bigskip\bigskip
}}
\vfill\eject
\pageno=2
\AB
We study the leading corrections to the emergent canonical commutation 
relations arising in the statistical mechanics of matrix models, by  
deriving several related Ward identities, and give conditions for these 
corrections to be small.  We show that emergent 
canonical commutators are possible only in matrix models 
in complex Hilbert space for which the numbers of fermionic and bosonic 
fundamental degrees of freedom are equal, suggesting that supersymmetry 
will play a crucial role.  Our results simplify, and sharpen, those obtained 
earlier by Adler and Millard.  
\AE
\bigskip\bigskip
\vfill\eject
\pageno=3
\centerline{\bf 1.~~Introduction}

It is widely believed that at distances of order the Planck length 
$\ell_P \sim 10^{-33}~{\rm cm}$ our conventional notions of the geometry 
of spacetime break down, as a result of quantum gravity effects.  One 
indication of the modifications in physics that might be expected is 
provided by string theory models of quantum gravity, 
in which several studies suggest a modification 
of the uncertainty relation of the form [1]
$$ \Delta x \Delta p \geq {\hbar \over 2} [1+\beta (\Delta p)^2+....],~~~
\beta >0~~~~~,\eqno(1a)$$
implying a finite minimum uncertainty $\Delta x_0=\hbar \beta^{1\over 2}$
in the vicinity of the Planck length.  As discussed by 
Kempf and collaborators [2], Eq.~(1a) corresponds to a correction to the 
Heisenberg canonical commutation relations of the form 
$$[x,p]=i\hbar(1+\beta p^2+...)~~~.\eqno(1b)$$

We wish in this paper to discuss modifications of the Heisenberg algebra 
arising in another context, that of the statistical mechanics of matrix 
models, and to compare them with Eqs.~(1a, b).  Several years ago,   
Adler proposed a set of rules for a generalized quantum or trace 
dynamics, which is a Lagrangian and Hamiltonian mechanics with arbitrary 
noncommutative phase space variables $q,p$, and this was developed in a 
series of papers with various collaborators [3].  
For theories in which the action is constructed as the trace 
of a sum of matrix products of $N \times N$ matrix variables, trace dynamics 
gives a powerful, basis independent, way of representing the same dynamics 
that can also be described in terms of the $N^2$ individual matrix elements.  
A significant new result emerging from this point of view was obtained by 
Adler 
and Millard [4], who argued that the {\it statistical mechanics} of trace 
dynamics takes the form of conventional quantum field theory, with 
the Heisenberg commutation relations holding for statistical averages 
over certain effective canonical variables obtained by projection from 
the original operator canonical variables.  Recently, it has become clear 
[5] that the underlying assumptions of trace dynamics are satisfied by 
matrix models, for which the methods of trace dynamics 
provide a very convenient calculational tool.  
Hence the results of Adler 
and Millard can be reinterpreted as providing a statistical mechanics of 
matrix models, and showing that thermal averages in this statistical 
mechanics can behave as Wightman functions in an emergent 
local quantum field theory.  These results, together with recent work [6]  
suggesting that the underlying dynamics for string theory may be a form 
of matrix model, raise
in turn the question of determining the form of the leading 
corrections to the Heisenberg algebra implied by the statistics of matrix  
models, formulating conditions for these corrections to be small, 
and seeing whether they can be related to the string theory result of 
Eqs.~(1a, 1b).   

An investigation of these questions is the focus of this 
paper, which is organized as follows.  In Sec.~2 we give a brief synopsis   
of the rules of trace dynamics in the context of matrix models.  We show 
that the conservation of the operator [4, 7] $\tilde C$ can be understood as 
a simple consequence of unitary invariance.  We also remark that,  
with Grassmann fermions, $\tilde C$ is independent of the classical 
parts of the matrix phase space variables, and review the statistical 
mechanics [4, 8] of matrix models.  In Sec.~3 we consider the simple case of 
a bosonic matrix model with Hamiltonian quadratic in the canonical momenta, 
and, making no approximations, derive a simplified form of 
the Ward identity used in Ref. [4] 
to obtain the effective canonical algebra.  
This analysis shows that there are 
corrections to the canonical commutator quadratic in the canonical momentum.
In Sec.~4 we use the symplectic formalism of Ref. [4] to repeat this 
calculation in the case of a general commutator/anticommutator of 
canonical variables in a generic matrix model,  
that can include fermions, and we generalize the treatment of [4] to  
allow nonzero sources for the classical parts of the matrix variables.     
From the analyses of Secs.~3 and 4, we formulate conditions 
for the corrections 
to the emergent canonical algebra to be small.  We show that these 
conditions require $\tilde C$ to be an intensive rather than extensive 
thermodynamic quantity, and that they can be satisfied in complex Hilbert 
space (if at all) only in matrix models with precisely equal numbers of 
bosonic and fermionic degrees of freedom.  This result strongly suggests 
that candidate matrix models for prequantum mechanics should be 
supersymmetric.  We conclude by generalizing the conditions 
to ones that permit the recovery of the full emergent quantum field theory 
structure derived in Ref. ~[4]. We also compare the prequantum corrections 
to the canonical algebra derived in 
Secs.~3 and 4, in which 
(as is usual in field theories) the spatial coordinate is simply a label, 
and the field variables are the dynamical canonical variables, to the   
string theory inspired expression of Eq.~(1b) in which $x$ is a coordinate 
operator.  

\bigskip
\centerline{{\bf 2.~~The Statistical Mechanics of Matrix Models}}

We begin by reviewing the  statistical mechanics of trace 
dynamics, taking into account the simplifications [5] that become 
possible when 
Grassmann algebras are employed to represent the fermion/boson distinction.  
Let $B_1$ and $B_2$ be two $N \times N$ matrices with matrix elements that 
are even grade elements of a complex Grassmann algebra, and Tr the ordinary 
matrix trace, which obeys the cyclic property 
$${\rm Tr} B_1 B_2 = \sum_{m,n}(B_1)_{mn}(B_2)_{nm} 
=\sum_{m,n} (B_2)_{nm}(B_1)_{mn}= {\rm Tr}B_2B_1~~~.
\eqno(2a)$$
Similarly, let $\chi_1$ and $\chi_2$ be two $N \times N$ matrices with 
matrix elements that are odd grade elements of a complex Grassmann algebra, 
which anticommute rather than commute, so that the cyclic property for these 
takes the form 
$${\rm Tr}\chi_1 \chi_2 = \sum_{m,n} (\chi_1)_{mn}(\chi_2)_{nm}
=-\sum_{m,n}(\chi_2)_{nm}(\chi_1)_{mn}=-{\rm Tr}\chi_2\chi_1~~~.
\eqno(2b)$$
The cyclic/anticyclic properties of Eqs.~(2a, 2b) are just those assumed  
for the trace operation {\bf Tr} of trace dynamics.\footnote{*}
{In Refs. [3, 4] 
the fermionic operators were realized as ordinary 
matrices with complex matrix 
elements, all of which anticommute with a grading operator $(-1)^F$ which 
formed part of the definition of the graded trace {\bf Tr}, for which 
fermions   
then obeyed Eq.~(2b) while bosons obeyed Eq.~(2a).
Since the use of Grassmann odd 
fermions eliminates the need for the inclusion of the $(-1)^F$ factor,
the graded trace obeying Eqs.~(2a, b) is here just the usual matrix 
trace, for which we use the customary notation Tr.} 
From Eqs.~(2a, b), one immediately derives the trilinear cyclic identities 
$$\eqalign{
{\rm Tr} B_1[B_2,B_3]=&{\rm Tr}B_2[B_3,B_1]={\rm Tr}B_3[B_1,B_2]~~~\cr
{\rm Tr} B_1\{B_2,B_3\}=&{\rm Tr}B_2\{B_3,B_1\}={\rm Tr}B_3\{B_1,B_2\}  \cr
{\rm Tr} B \{\chi_1,\chi_2\}=&{\rm Tr}\chi_1 [\chi_2,B]
={\rm Tr} \chi_2[\chi_1,B]  \cr
{\rm Tr} \chi_1\{B,\chi_2\}=&{\rm Tr}\{\chi_1,B\}\chi_2
={\rm Tr}[\chi_1,\chi_2]B ~~~, \cr
}\eqno(2c) $$
which are used repeatedly in trace dynamics calculations.   

The basic observation of trace dynamics is that given the trace of a 
polynomial $P$ constructed from noncommuting matrix or operator variables (we 
shall use the terms ``matrix'' and ``operator'' interchangeably in the 
following discussion), 
one can define a derivative of the $c$-number ${\rm Tr} P$ with respect to 
an operator variable ${\cal O}$ by varying and then cyclically permuting 
so that in each term the factor $\delta {\cal O}$ stands on the right, 
giving the fundamental definition 
$$\delta {\rm Tr P}={\rm Tr} {\delta {\rm Tr}P \over \delta {\cal O}} 
\delta {\cal O}~~~,\eqno(3a)$$
or in the condensed notation that we shall use throughout this paper, in 
which ${\bf P} \equiv {\rm Tr}P$, 
$$\delta {\bf P}   = {\rm Tr} {\delta {\bf P} \over \delta {\cal O}}
\delta {\cal O}~~~.\eqno(3b)$$
Letting ${\bf L}[\{q_r\},\{\dot q_r\}]$ be a trace Lagrangian that is a 
function of the bosonic or fermionic operators $\{q_r\}$ and their time 
derivatives (which are all assumed to obey the cyclic relations of 
Eqs.~(2a-c) under the trace), and requiring that the trace 
action ${\bf S}=\int dt {\bf L}$ be stationary with respect to variations 
of the $q_r$'s that preserve their bosonic or fermionic type, 
one finds [3] the operator 
Euler-Lagrange equations 
$${\delta {\bf L} \over \delta q_r} -
{d \over dt} {\delta {\bf L} \over \delta \dot q_r} =0~~~.\eqno(3c)$$
Because, by the definition of Eq.~(3a), we have 
$$\left( {\delta {\bf L} \over \delta q_r} \right)_{ij}=
{\partial {\bf L} \over \partial (q_r)_{ji} }~~~,\eqno(3d)$$
for each $r$ the single Euler-Lagrange equation of Eq.~(3c) is equivalent 
to the $N^2$ Euler-Lagrange equations obtained by regarding ${\bf L}$ 
as a function of the $N^2$ matrix element variables $(q_r)_{ji}$.  
Defining the momentum operator $p_r$ conjugate to $q_r$, which is of the 
same bosonic or fermionic type as $q_r$, by 
$$p_r \equiv {\delta {\bf L} \over \delta \dot q_r}~~~,\eqno(4a)$$
the trace Hamiltonian {\bf H} is defined by 
$${\bf H}={\rm Tr}\sum_rp_r \dot q_r - {\bf L}~~~.\eqno(4b)$$
In correspondence with Eq.~(3d), the matrix elements $(p_r)_{ij}$ of the 
momentum operator $p_r$ just correspond to the momenta canonical to the 
matrix element variables $(q_r)_{ji}$.  
Performing general same-type operator variations, and using the 
Euler-Lagrange 
equations, we find from Eq.~(4b) that the trace Hamiltonian {\bf H} is a 
trace functional of the operators $\{q_r\}$ and $\{p_r\}$, 
$${\bf H}= {\bf H}[\{q_r\},\{p_r\}]~~~,\eqno(5a)$$
with the operator derivatives 
$${\delta {\bf H} \over \delta q_r}=-\dot p_r~,~~~
{\delta {\bf H} \over \delta p_r}=\epsilon_r \dot q_r~,~~~\eqno(5b)$$
with $\epsilon_r=1(-1)$ according to whether $q_r,p_r$ are 
bosonic (fermionic).  
Letting {\bf A} and {\bf B} be two trace functions of the operators 
$\{q_r\}$ and $\{p_r\}$, it is convenient to define the {\it generalized 
Poisson bracket} 
$$\{{\bf A}, {\bf B} \}={\rm Tr} \sum_r \epsilon_r \left(
{\delta {\bf A} \over \delta q_r}{\delta {\bf B} \over \delta p_r}
-{\delta {\bf B} \over \delta q_r} {\delta {\bf A} \over \delta p_r} \right)
~~~.\eqno(6a)$$
Then using the Hamiltonian form of the equations of motion, one readily 
finds that for a general trace functional ${\bf A}[\{q_r\},\{p_r\}]$, 
the time derivative is given by 
$${d \over dt} {\bf A}=\{ {\bf A}, {\bf H} \}~~~;\eqno(6b)$$
in particular, letting {\bf A} be the trace Hamiltonian {\bf H}, and using 
the fact that the generalized Poisson bracket is antisymmetric in its 
arguments, it follows that the time derivative of {\bf H} vanishes. 
An important property of the generalized Poisson bracket is that it 
satisfies [3] the Jacobi identity, 
$$\{ {\bf A},\{ {\bf B},{\bf C} \}\}    
+\{ {\bf C},\{ {\bf A},{\bf B} \}\}    
+\{ {\bf B},\{ {\bf C},{\bf A} \}\} =0~~~.\eqno(6c)$$   
As a consequence, if ${\bf Q_1}$ and ${\bf Q_2}$ are two conserved charges, 
that is if
$$0={d \over dt} {\bf Q_1} = \{ {\bf Q_1}, {\bf H} \}~,~~~ 
  0={d \over dt} {\bf Q_2} = \{ {\bf Q_2}, {\bf H} \}~,~~~ 
  \eqno(6d)$$ 
then their generalized Poisson bracket $\{ {\bf Q_1}, {\bf Q_2} \}$ 
also has a vanishing generalized Poisson bracket with {\bf H}, and is 
conserved.  This has the consequence that Lie algebras of symmetries can  
be represented as Lie algebras of trace functions under the generalized 
Poisson bracket operation.  

A significant feature of trace dynamics is that, as discovered by 
Millard [7], the anti-self-adjoint operator [7, 4] 
$$\tilde C \equiv \sum_{r~{\rm bosons}}[q_r,p_r]-\sum_{r~{\rm fermions}}
\{q_r, p_r\}~~~\eqno(7)$$
is conserved by the dynamics.  As we shall now show, conservation of  
$\tilde C$ holds whenever the trace dynamics has a global unitary invariance, 
that is, whenever the trace Hamiltonian obeys
$${\bf H}[\{ U^{\dagger}q_rU\},\{U^{\dagger}p_rU\}]={\bf H}[\{q_r\},\{p_r\}]
~~~\eqno(8a)$$
for a constant unitary $N \times N$ matrix $U$, or equivalently, by 
Eq.~(4b), whenever the trace Lagrangian obeys 
$${\bf L}[\{U^{\dagger} q_r U\},\{U^{\dagger}\dot q_r U\}]=
{\bf L}[\{q_r\},\{\dot q_r\}]~~~.\eqno(8b)$$
Letting $U=\exp\Lambda$, with $\Lambda$ an anti-self-adjoint bosonic 
generator matrix, 
and expanding to first order in $\Lambda$, Eq.~(8a) implies that 
$${\bf H}[\{q_r -[\Lambda,q_r]\}, \{p_r - [\Lambda,p_r]\}]
={\bf H}[\{q_r\},\{p_r\}] ~~~.\eqno(9a)$$
But applying the definition of the variation of a trace functional given 
in Eq.~(3b), Eq.~(9a) becomes 
$${\rm Tr} \sum_r \left(- {\delta {\bf H} \over \delta q_r}
[\Lambda,q_r] - {\delta {\bf H} \over \delta p_r} [\Lambda,p_r] \right) 
=0~~~,\eqno(9b)$$
which by use of the trilinear cyclic identities of Eq.~(2c) yields  
$${\rm Tr}  \Lambda 
\sum_r \left( 
{\delta {\bf H} \over \delta q_r}  q_r -\epsilon_r q_r {\delta {\bf H} 
\over \delta q_r} 
+{\delta {\bf H} \over \delta p_r}  p_r -\epsilon_r p_r {\delta {\bf H} 
\over \delta p_r} \right)=0.~~~~\eqno(9c)$$
Since the generator $\Lambda$ is an arbitrary anti-self-adjoint $N \times N$ 
matrix, the anti-self-adjoint matrix that multiplies it in Eq.~(9c) must 
vanish, giving the matrix  identity 
$$ 
\sum_r \left( 
{\delta {\bf H} \over \delta q_r}  q_r-\epsilon_r q_r {\delta {\bf H} 
\over \delta q_r} 
+{\delta {\bf H} \over \delta p_r}  p_r -\epsilon_r p_r {\delta {\bf H} 
\over \delta p_r} \right)=0.~~~~\eqno(10a)$$
But now substituting the Hamilton equations of Eq.~(5b), Eq.~(10a) takes 
the form 
$$\eqalign{
0=&\sum_r \left( 
-\dot p_r  q_r +\epsilon_r q_r \dot p_r 
+\epsilon_r \dot q_r p_r - p_r \dot q_r \right)  \cr
=&{d \over dt}\sum_r  \left( -p_r q_r + \epsilon_r q_r p_r \right)   \cr
=& {d \over dt}\left( \sum_{r~{\rm bosons}}[q_r,p_r]-\sum_{r~{\rm fermions}}
\{q_r, p_r\} \right)~~~,\cr 
}\eqno(10b)$$
completing the demonstration of the conservation of $\tilde C$.  

Corresponding to the fact that $\tilde C$ is conserved in any matrix model 
with a global unitary invariance, it is easy to see [4, 8] that 
$\tilde C$ can be used to construct the generator of global unitary 
transformations of the Hilbert space basis.  Consider the trace functional  
$${\bf G}_{\Lambda} =-{\rm Tr} \Lambda \tilde C~~~,\eqno(11a)$$
with $\Lambda$ a fixed bosonic anti-self-adjoint operator, which can be 
rewritten, using cyclic 
invariance of the trace, as 
$${\bf G}_{\Lambda}={\rm Tr}\sum_r[\Lambda,p_r]q_r 
=-{\rm Tr}\sum_r p_r [\Lambda ,q_r]~~~.\eqno(11b)$$
Hence for the variations of $p_r$ and $q_r$ induced 
by ${\bf G}_{\Lambda}$ as 
canonical generator, which have a structure analogous to the 
Hamilton equations of Eq.~(5b), we get     
$$\eqalign{
\delta p_r=&-{\delta {\bf G}_{\Lambda} \over \delta q_r} 
=-[\Lambda,p_r]~~~,\cr
\delta q_r=&\epsilon_r {\delta {\bf G}_{\Lambda} \over \delta p_r} 
=-[\Lambda,q_r]~~~.\cr 
}\eqno(11c)$$ 
Comparing with Eqs.~(8a) and (9a), we see that these have just the form of an 
infinitesimal global unitary transformation.  

For each phase space variable $q_r,p_r$, let us define the {\it classical 
part} $q_r^c, p_r^c$ and the noncommutative remainder $q_r^{\prime}, 
p_r^{\prime}$, by
$$\eqalign{
q_r^c=&{1 \over N}{\rm Tr} q_r~~~~~p_r^c={1 \over N} {\rm Tr} p_r~~~,\cr
q_r^{\prime}=&q_r-q_r^c~~~~~p_r^{\prime}=p_r-p_r^c~~~~,\cr
}\eqno(12a)$$
so that bosonic $q_r^c,p_r^c$ are $c$-numbers, fermionic $q_r^c,p_r^c$ 
are Grassmann $c$-numbers (where by a $c$-number we mean a multiple of $1_N$,
the $N \times N$ unit matrix), and the remainders are traceless,
$${\rm Tr} q_r^{\prime}={\rm Tr}p_r^{\prime}=0~~~.\eqno(12b)$$
Then since $q_r^c,p_r^c$ commute (anticommute) with $q_s^{\prime},
p_s^{\prime}$ for $r,s$ both bosonic (fermionic), we see that the classical 
parts of the phase space variables make no contribution to $\tilde C$, and 
Eq.~(7) can be rewritten as 
$$\tilde C=\sum_{r~{\rm bosons}}[q_r^{\prime},p_r^{\prime}]
-\sum_{r {\rm~fermons}}\{q_r^{\prime},p_r^{\prime}\}~~~.\eqno(12c)$$
Thus $\tilde C$ is completely independent of the values of the classical 
parts of the matrix phase space variables.

Making the assumption that trace dynamics is ergodic 
(which undoubtedly requires an interacting as opposed to a free theory, 
and may presuppose taking the $N \rightarrow \infty $ limit), 
one can then analyze [4] the statistical mechanics of trace dynamics for the 
generic case in which the conserved quantities are the trace Hamiltonian 
{\bf H} and the operator $\tilde C$.  As discussed in detail in the 
second paper cited in Ref. [5], 
the analysis of [4] carries over to the case in which the fermions are 
represented by Grassmann matrices; the demonstration of a generalized 
Liouville theorem still holds, and the requirements for convergence of 
the partition function are much less stringent, eliminating the complexities 
addressed in Appendix F of Ref. [4].  With Grassmann fermions,  
for the typical models we are studying the 
bosonic part of $H$ is a positive operator, from which  
{\bf H}  inherits good positivity properties .   
The canonical ensemble then takes the simple form given 
in Eq.~(48c) of [4], 
$$\eqalign{
\rho=&Z^{-1} \exp(-\tau {\bf H} -{\rm Tr} \tilde \lambda \tilde C)    \cr
Z=&\int d\mu \exp(-\tau {\bf H} -{\rm Tr} \tilde \lambda \tilde C )~~~,\cr
}\eqno(13)$$
with $d \mu$  the invariant matrix (or operator) phase space measure 
provided by Liouville's theorem, with $\tau$ a real number, and with 
$\tilde \lambda$ an anti-self-adjoint matrix  that in the 
generic case (which we assume) has no zero eigenvalues. 
(Equation (13) can be derived directly [4] by 
maximizing the entropy subject to the constraints imposed by the conservation 
of {\bf H} and $\tilde C$, or indirectly [8] by first calculating the  
corresponding microcanonical ensemble corresponding to these conserved 
quantities, and then using standard statistical physics methods to calculate 
the canonical ensemble from the microcanonical one.)   We wish to make 
two points about the partition function defined in Eq.~(13).  First of all, 
it is not invariant under the unitary transformation of Eq.~(8a) for fixed 
$\tilde \lambda$, but is invariant when $\tilde \lambda$ is simultaneously 
transformed to $U^{\dagger} \tilde \lambda U$; hence the partition function 
breaks unitary invariance, but has a specific form of unitary covariance.  
Second, the partition function contains a weighted sum over all possible 
commutators $[q_r,p_s]$ for bosonic variables and all possible 
anticommutators $\{q_r,p_s\}$ for fermionic variables; there is no 
restriction to the classical or quantum mechanical evaluation of these 
commutators/anticommutators as 0 or $i\delta_{rs}$ respectively. 
However, statistical integrals like Eq.~(13) are typically dominated by 
specific regions of the integration domain, and we will see, by a 
study of the Ward identities following from Eq.~(13), that this can lead 
to {\it effective} quantum mechanical commutators inside statistical 
averages.  
The structure of the Ward identities or equipartition theorems following from 
Eq.~(13)
will be reanalyzed in the next two sections without making 
approximations used in [4], so as to determine the leading corrections 
to the emergent canonical commutation relations. From this analysis we 
will infer a set of conditions for obtaining the full emergent quantum field 
theory structure of [4].  
\vfill
\eject
\bigskip    
\centerline{\bf 3.~~Corrections to the Bosonic Commutator $[q_s,p_r]$ in }
\centerline{\bf a Simplified Unitary Invariant Matrix Model}

We consider in this section the simplified bosonic matrix model with 
trace Hamiltonian
$${\bf H}={\rm Tr}\left[ \sum_r {1 \over 2} p_r^2 + V(\{q_r\}) \right]
~~~,\eqno(14a)$$
with the $q_r$ self-adjoint $N \times N$ complex matrix variables and with 
$V$ a global unitary invariant potential.  This form is general enough to 
include the matrix model forms of the bosonic field theories of greatest 
interest, including the Goldstone model, non-Abelian gauge models, and the 
Higgs model.  As we saw in the previous section, the Hamiltonian dynamics 
for this model conserves both the real 
number {\bf H} and the matrix $\tilde C$, 
which in this case is given simply by
$$\tilde C=\sum_r[q_r,p_r]~~~.\eqno(14b)$$
Letting $\rho$ and $Z$ be respectively the canonical ensemble and partition 
function given in terms of {\bf H} and $\tilde C$ by Eq.~(13), we define 
the ensemble average of an arbitrary function $\cal O$ of the dynamical 
variables by
$$\langle {\cal O} \rangle_{AV}=\int d\mu \rho {\cal O} = 
Z^{-1}\int d\mu e^{-\tau {\bf H}-{\rm Tr}(\tilde \lambda \tilde C)} {\cal O}
~~~.\eqno(15)$$
Letting $\cal O$ be the conserved operator $\tilde C$, and noting that 
the right hand side of Eq.~(13) can be a function only of the ensemble 
parameters $\tilde \lambda$ and $\tau$, we have 
$$\langle \tilde C \rangle_{AV}=f(\tilde \lambda,\tau),~~~\eqno(16a)$$
with $f$ an anti-self-adjoint matrix, which in general can be written    
as a phase matrix $i_{\rm eff}$ times a commuting magnitude matrix $|f|$, 
$$f=i_{\rm eff}|f|,~~~ i_{\rm eff}^2=-1,~~~i_{\rm eff}^{\dagger}=
-i_{\rm eff},
~~~[i_{\rm eff},|f|~]=0~~~.\eqno(16b)$$

We shall now specialize to an ensemble for which the magnitude matrix $|f|$ 
makes no distinction among the different bases in Hilbert space, and so 
takes the form of a positive real multiple (which we shall call $\hbar$) of 
the unit matrix.  
[As discussed in Appendix B of Ref. [4], when ${\bf H}$ can be expressed 
in terms of the phase space operators $\{ q_r,p_r \}$ using only real 
number coefficients, this assumption implies that we are restricting 
attention 
to the special class of ensembles for which 
$\tilde \lambda=i_{\rm eff} \lambda$, with $\lambda$ 
a real multiple of the unit matrix.]
Equations (16a, b) then become  
$$\langle \tilde C \rangle_{AV}=i_{\rm eff} \hbar~~~.\eqno(16c)$$
Since for finite $N$ we necessarily have ${\rm Tr} \tilde C=0$, the 
phase matrix $i_{\rm eff}$ must have vanishing trace, 
$${\rm Tr} i_{\rm eff}=0~~~,\eqno(16d)$$
which implies that $i_{\rm eff}$ has $N/2$ eigenvalues $i$  and $N/2$ 
eigenvalues $-i$.  Thus, we are making a choice of ensemble for which 
the $U(N)$ symmetry of ${\bf H}$ is broken, by the term 
${\rm Tr} \tilde \lambda \tilde C$, to $U(N/2) \times U(N/2) \times R$, 
with $R$ the discrete reflection symmetry that interchanges the eigenvalues 
$\pm i$ of $i_{\rm eff}$.  This is clearly the largest symmetry group of 
the ensemble for which one can have $\langle \tilde C \rangle_{AV} \not=0$; 
if one were to attempt to preserve the full $U(N)$ symmetry by  
taking an ensemble 
with $\tilde \lambda=i\lambda$, with $\lambda$ a $c$-number, 
then in the canonical ensemble the term 
${\rm Tr}\tilde \lambda \tilde C$ would vanish by virtue of the tracelessness 
of $\tilde C$, and the resulting ensemble would have 
$\langle \tilde C \rangle_{AV} =0$. Requiring the largest possible nontrivial  
symmetry group plays the role in our derivation of giving a single Planck 
constant for all pairs of canonical variables; if on the other hand, we 
were to sacrifice all of the $U(N)$ symmetry by allowing generic $\tilde 
\lambda$, then the emergent canonical commutation relations derived below 
would generically yield $N/2$ different $\hbar$'s for the $N/2$ pairs 
of canonical variables.  It would clearly be desirable to have a deeper 
justification from first principles of our choice of ensemble, 
perhaps based on a more detailed understanding of the underlying dynamics, 
but at present we must simply introduce it as a postulate.  

For this choice of ensemble, let us now consider the Ward identity 
obtained from   
$$Z \langle {\rm Tr} \tilde C p_r \rangle_{AV}
=\int d\mu e^{-\tau {\bf H}-{\rm Tr} \tilde \lambda \tilde C}{\rm Tr} 
\tilde C p_r, ~~~\eqno(17a)$$
by using invariance of the measure $d\mu$ under a constant shift 
of $p_s$, which implies 
$$\eqalign{
0=&\int d\mu \delta_{p_s}\left[ 
e^{-\tau {\bf H}-{\rm Tr} \tilde \lambda \tilde C}  
{\rm Tr}   \tilde C p_r \right]  \cr
=&\int d\mu e^{-\tau {\bf H}-{\rm Tr} \tilde \lambda \tilde C}  \cr
\times&
\left[(-\tau \delta_{p_s}{\bf H}-{\rm Tr} \tilde \lambda 
\delta_{p_s} \tilde C ) {\rm Tr}   \tilde C p_r
+{\rm Tr} (\delta_{p_s} \tilde C) p_r 
+{\rm Tr}\tilde C  \delta_{rs} \delta p_s \right]~~~. \cr
}\eqno (17b)$$
Now from Eqs.~(14a, b) we have 
$$\delta_{p_s}{\bf H}={\rm Tr} p_s \delta p_s,~~~
\delta_{p_s} \tilde C=[q_s,\delta p_s]~~~.\eqno(18)$$
Substituting these into Eq.~(17b), multiplying by $Z^{-1}$, and using 
the trilinear cyclic identities of Eq.~(2c), we get 
$$0=\langle (-\tau {\rm Tr} p_s \delta p_s-{\rm Tr}[\tilde \lambda,q_s]
\delta p_s )  {\rm Tr}   \tilde C p_r
+{\rm Tr}[p_r,q_s] \delta p_s + {\rm Tr} \tilde C 
\delta_{rs} \delta p_s  \rangle_{AV} ~~~,\eqno(19a)$$
which since $\delta p_s$ is an arbitrary self-adjoint matrix, implies that 
the operator multiplying $\delta p_s$ inside the trace must 
vanish, 
$$0=\langle (-\tau  p_s -[\tilde \lambda,q_s]) {\rm Tr} \tilde C p_r 
+[p_r,q_s]  + \tilde C  
\delta_{rs}  \rangle_{AV} ~~~.\eqno(19b)$$
Since $\tilde \lambda$ is a constant matrix, it can be taken outside 
the ensemble average, and so the second term in Eq.~(19b) takes the form 
$$-[\tilde \lambda, \langle q_s {\rm Tr} \tilde C p_r \rangle_{AV} ]
~~~,\eqno(19c)$$ 
which vanishes since the ensemble average inside the commutator in Eq.~(19c) 
is a matrix function only of $\tilde \lambda$ and $\tau$, and hence commutes 
with $\tilde \lambda$.  
Substituting $p_r=p_r^c+p_r^{\prime}$ into the term multiplied by $\tau$ 
in Eq.~(19b), and using ${\rm Tr}\tilde C p_r^c$$=p_r^c {\rm Tr}\tilde C=0$, 
the traceless part of Eq.~(19b) reduces to 
$$0=\langle -\tau p_s^{\prime} {\rm Tr} \tilde C p_r^{\prime} +[p_r,q_s] 
+\tilde C \delta_{rs}\rangle_{AV}~~~.\eqno(20a)$$
Since $\langle \tilde C \rangle_{AV} = i_{\rm eff} \hbar$, this 
equation can be rewritten as the {\it exact} relation 
$$\langle [q_s,p_r] \rangle_{AV}=i_{\rm eff}\hbar \delta_{rs}
-\tau \langle p_s^{\prime} {\rm Tr} \tilde C p_r^{\prime} \rangle_{AV}~~~ 
,\eqno(20b)$$
showing that the ensemble averaged commutator of the canonical 
coordinate and momentum operators has the form of the usual quantum 
mechanical canonical commutator, with $i_{\rm eff}$ playing the role 
of the imaginary unit, and with a correction term proportional 
to $\tau $ that is quadratic in the non-classical parts of the 
canonical momenta. Using Eqs.~(14b) and (16c), Eq.~(20b) can also be 
written in the form
$$\sum_{t,u}\langle [q_t,p_u]\rangle_{AV}(\delta_{tr}\delta_{us}
-\delta_{tu}\delta_{rs})=
-\tau \langle p_s^{\prime} {\rm Tr} \tilde C p_r^{\prime} \rangle_{AV}~~~ 
.\eqno(20c)$$
The derivation given here sharpens that given in Adler and Millard [4], 
both in 
that here no approximations have been made, and that because we are 
working in complex Hilbert 
space, we have not had to first project out the parts of $q_s$ and 
$p_r$ that commute with $i_{\rm eff}$.\footnote{*}{As discussed in the 
Appendix, this projection is needed in real and quaternionic Hilbert 
space.}

We conclude this section with several remarks on the principal result of 
Eq.~(20b).  First of all, the fluctuations about the ensemble average 
are fundamental to the possibility of emergent quantum behavior.  The
average of the commutator on the left hand side of Eq.~(20b) is not 
the same as the commutator of the averages, which in fact vanishes since 
$\langle q_s\rangle_{AV}$ and $\langle p_r \rangle_{AV}$ are both functions 
only of $\tau$ and $\tilde \lambda$,
$$\langle [q_s,p_r]\rangle_{AV} \not = [\langle q_s \rangle_{AV}, 
\langle p_r \rangle_{AV}]=0~~~.\eqno(21a)$$
Secondly, in order to have emergent quantum behavior, the dynamics must 
be such that 
the second term on the right hand side  of Eq.~(20b) is much smaller 
than the first term on the right hand side, that is, one must have 
$$\hbar^{-1}\tau|\langle p_s^{\prime} {\rm Tr} \tilde C p_r^{\prime} 
\rangle_{AV} | <<1~~~.\eqno(21b)$$
We shall now show that this condition cannot be satisfied if $\tilde C$ is 
an extensive thermodynamic quantity that grows linearly with the size 
of the system.  To see this, we note that a second Ward identity, similar 
in form to Eq.~(20b), can be derived by starting from 
$$0=\int d\mu \delta_{p_s}\left[ 
e^{-\tau {\bf H}-{\rm Tr} \tilde \lambda \tilde C}  
{\rm Tr}  i_{\rm eff} \hbar  p_r \right]  ~~~, \eqno (22a)$$
in which the factor of $\tilde C$ multiplying $p_r$ has been replaced by 
its ensemble average $i_{\rm eff} \hbar$, and then proceeding as in 
Eqs.~(17b-20b) 
above.  The resulting Ward identity is 
$$0=i_{\rm eff}\hbar \delta_{rs}
-\tau \langle p_s^{\prime} {\rm Tr} i_{\rm eff}\hbar p_r^{\prime} 
\rangle_{AV}~~~
,\eqno(22b)$$
and is an analog in our context of the usual equipartition theorem of 
classical statistical mechanics.  Now if $\tilde C$ were an extensive 
quantity, the difference between $\tilde C$ and its ensemble average 
$i_{\rm eff} \hbar$ would be a fluctuation that vanishes as the system size 
becomes infinite, which would make it permissible to accurately approximate 
the right hand side of Eq.~(20b) by the right hand side of Eq. (22b).  
This would lead to the conclusion $\langle[q_s,p_r]\rangle_{AV}=0$, that 
is, the thermodynamics would give emergent classical, rather than quantum 
mechanical, behavior.  

Thus, the inequality of Eq.~(21b) can hold only if 
$\tilde C$ is not extensive; this conclusion is consistent with the 
observation  that to have the average of $\tilde C$ play 
the role of the intensive quantity $i_{\rm eff} \hbar$, one would expect that 
$\tilde C$ should behave as a thermodynamically intensive quantity.  
We shall give further evidence for this conclusion in the next section, 
where we consider systems containing fermions as well as bosons.  
 
Although we have first discussed the purely bosonic model of this section 
for expository reasons, it is easy to see from Eqs.~(20b, c) that the 
inequality of Eq.~(21b) cannot hold in a purely bosonic system.  When the 
right hand side of Eq.~(20c) can be neglected, multiplying by $\delta_{rs}$  
and summing over $r,s$ gives the relation  
$$\sum_{t,u}\langle [q_t,p_u]\rangle_{AV} \delta_{tu} (1-N)=0~~~
,\eqno(22c)$$
which for $N>1$ implies that $\langle[q_s,p_r]\rangle_{AV}=0$, again 
giving classical behavior.  Thus, a purely bosonic matrix dynamics 
system in complex Hilbert space cannot have emergent quantum behavior.   
However, we shall see in the 
next section that in the interesting case in which the numbers of bosonic 
and fermionic degrees of freedom are equal, a condition that holds for  
supersymmetric theories, the relation of Eq.~(22c) is modified, and 
emergent quantum behavior becomes possible.

\bigskip
\centerline{\bf 4.~~Corrections to the Full Canonical Algebra in a }
\centerline{\bf General Unitary Invariant Matrix Model  with 
Classical Sources}

Although one can give derivations similar to that of the previous 
section for other canonical commutators (e.g., the $[p_r,p_s]$ and 
$[q_r,q_s]$ commutators, or the $[p_s,q_r]$ commutator obtained by 
interchanging the roles of $p$ and $q$ in the above derivation), and 
their fermionic anticommutator analogs, it is most 
efficient in the general case to use the symplectic formalism introduced 
by Adler and Millard in Ref. [4].   In this notation one defines 
$x_1=q_1,x_2=p_1,x_3=q_2,x_4=p_2,...,x_{2R-1}=q_R,x_{2R}=p_R$ for the 
matrix phase space variables; in terms of these, the Hamilton equations 
of Eq.~(5b), the generalized Poisson bracket of Eq.~(6a), and the  
conserved operator $\tilde C$ of Eq.~(7) take the form 
$$\eqalign{
\dot x_r=&\sum_{s=1}^{2R} \omega_{rs} {\delta {\bf H} \over \delta x_s}  \cr
\{{\bf A},{\bf B}\}=&{\rm Tr}\sum_{r,s=1}^{2R}
{\delta {\bf A} \over \delta x_r}  
\omega_{rs}{\delta {\bf B}\over \delta x_s} \cr
\tilde C=&\sum_{r,s} x_r \omega_{rs} x_s ~~~,\cr
}\eqno(23a)$$
with $\omega$ a numerical symplectic metric given (in terms of standard 
Pauli matrices $\tau_{1,2,3}$)
by $i\tau_2$ for a bosonic pair of canonical variables, and by 
$-\tau_1$ for a fermionic canonical pair (see Eqs. (10a-c) on p. 202, 
and Eqs. (10$a^{\prime}$-$c^{\prime}$) on p. 224, of Ref.~[4].)  This 
symplectic metric obeys [4] the useful identities 
$$\eqalign{
\omega_{sr}=&-\epsilon_r\omega_{rs}=-\epsilon_s\omega_{rs} \cr
\sum_r \omega_{rs}\omega_{rt}=&\sum_r \omega_{sr} \omega_{tr}
=\delta_{st}~~~.\cr
}\eqno(23b)$$

We shall now consider a matrix dynamics generated by a general trace 
Hamiltonian ${\bf H}$, that can contain fermionic as well as bosonic 
degrees of freedom, and in the statistical partition function shall allow the 
presence of nonvanishing classical sources $J_r^c$ for the classical 
parts $x_r^c$ of the 
phase space variables [cf. Eqs.~(12a-c) above.]  Thus we start from the 
ensemble 
$$\eqalign{
\rho=&Z^{-1} \exp(-\tau {\bf H} -{\rm Tr} \tilde \lambda \tilde C-\sum_r 
J_r^c x_r^c)    \cr
Z=&\int d\mu \exp(-\tau {\bf H} -{\rm Tr} \tilde \lambda \tilde C-\sum_r 
J_r^c x_r^c)~~~,\cr
}\eqno(24)$$
which is now used in place of the ensemble of Eq.~(13) in the 
definition of $\langle {\cal O} \rangle_{AV}$ given in Eq.~(15).  For 
this ensemble, we consider the Ward identity obtained 
by using shift invariance of the integration measure $d \mu$ starting from 
$$Z \langle {\rm Tr} \tilde C \sigma_t x_t^{\prime} \rangle_{AV}
=\int d\mu e^{-\tau {\bf H}-{\rm Tr} \tilde \lambda \tilde C -
\sum_r J_r^c x_r^c }{\rm Tr} 
\tilde C \sigma_t x_t^{\prime}, ~~~\eqno(25a)$$
where $\sigma_t$ are a set of $c$-number auxiliary parameters that are 
complex for bosonic $x_t$, and complex Grassmann for fermionic $x_t$.  
As in Eqs.~(12a, b) above, we use the notation $x_t^{\prime}$ to denote 
the noncommutative part of $x_t$ that remains when the classical part 
is subtracted away, that is, $x_t^{\prime}=x_t-x_t^c$.

The Ward identity derivation now proceeds exactly as in Eqs.~(17b-20b).  
Making a constant shift of the noncommutative part $x_s^{\prime}$ of the 
phase space variable $x_s$, we have 
$$\eqalign{
0=&\int d\mu \delta_{x_s^{\prime}}\left[ 
e^{-\tau {\bf H}-{\rm Tr} \tilde \lambda \tilde C -
\sum_r J_r^c x_r^c}
{\rm Tr}   \tilde C \sigma_t x_t^{\prime} \right]  \cr
=&\int d\mu e^{-\tau {\bf H}-{\rm Tr} \tilde \lambda \tilde C -
\sum_r J_r^c x_r^c}  \cr
\times&
\left[(-\tau \delta_{x_s^{\prime}}{\bf H}-{\rm Tr} \tilde \lambda 
\delta_{x_s^{\prime}} \tilde C ) {\rm Tr}   \tilde C \sigma_t x_t^{\prime}
+{\rm Tr} (\delta_{x_s^{\prime}} \tilde C) \sigma_t x_t^{\prime} 
+{\rm Tr}\tilde C \sigma_t  \delta_{st} \delta x_s^{\prime} \right]~~~. \cr
}\eqno (25b)$$
Now from Eqs.~(23a, b) we have 
$$\eqalign{
\delta_{x_s^{\prime}}{\bf H}=&{\rm Tr} \left({\delta {\bf H} \over   
\delta x_s^{\prime}}\right)^{\prime}\delta x_s^{\prime}
={\rm Tr} \sum_r {\dot x}_r^{\prime}\omega_{rs} \delta x_s^{\prime}~~~,\cr
{\rm Tr} \tilde \lambda \delta_{x_s^{\prime}} \tilde C=&
{\rm Tr}\tilde \lambda \sum_r \omega_{rs} (x_r^{\prime} \delta x_s^{\prime} 
-\epsilon_r \delta x_s^{\prime} x_r^{\prime})=
{\rm Tr} [\tilde \lambda,\sum_r \omega_{rs} x_r^{\prime}] \delta x_s^{\prime}
~~~,\cr
{\rm Tr} (\delta_{x_s^{\prime}} \tilde C) \sigma_t x_t^{\prime}=&  
{\rm Tr} [\sigma_t x_t^{\prime},\sum_r \omega_{rs} x_r^{\prime}] 
\delta x_s^{\prime}~~~.\cr
}\eqno(26)$$
Substituting these into Eq.~(25b) and multiplying by $Z^{-1}$, we get 
$$\eqalign{
0=&\langle (-\tau {\rm Tr} \sum_r {\dot x}_r^{\prime} \omega_{rs} 
\delta x_s^{\prime}-{\rm Tr}[\tilde \lambda,\sum_r \omega_{rs} x_r^{\prime}]
\delta x_s^{\prime} )  {\rm Tr}   \tilde C \sigma_t x_t^{\prime}\cr
+&{\rm Tr}[\sigma_t x_t^{\prime},\sum_r \omega_{rs} x_r^{\prime}] 
\delta x_s^{\prime} 
+ {\rm Tr} \tilde C \sigma_t
\delta_{st} \delta x_s^{\prime}  \rangle_{AV} ~~~,\cr
}\eqno(27a)$$
which since $\delta x_s^{\prime}$ is an arbitrary traceless 
matrix (with the adjointness properties of $x_s^{\prime}$), implies that 
the traceless part of the operator multiplying 
$\delta x_s^{\prime}$ inside the trace must 
vanish, 
$$\eqalign{
0=&\langle (-\tau \sum_r {\dot x}_r^{\prime} \omega_{rs} 
-[\tilde \lambda,\sum_r \omega_{rs} x_r^{\prime}]) 
{\rm Tr} \tilde C \sigma_t x_t^{\prime} \cr
+&[\sigma_t x_t^{\prime},\sum_r \omega_{rs} x_r^{\prime}]  
+ \tilde C \sigma_t \delta_{st}  \rangle_{AV} ~~~.\cr
}\eqno(27b)$$
Since $\tilde \lambda$ is a constant matrix, as before it can be taken 
outside 
the ensemble average, and so the second term in Eq.~(27b) takes the form 
$$-[\tilde \lambda, \sum_r \omega_{rs} \langle x_r^{\prime} {\rm Tr} 
\tilde C \sigma_t x_t^{\prime} \rangle_{AV} ]
~~~,\eqno(27c)$$ 
which again vanishes since the ensemble average inside the commutator 
is a matrix function only of $\tilde \lambda$ and $\tau$.    Contracting 
the remainder of Eq.~(27b) with $\sum_s\omega_{us}$, using Eqs.~(16c) and 
(23b), and noting that $[x_u^{\prime},\sigma_t x_t^{\prime}]$
$=[x_u,\sigma_t x_t]$ because the classical parts do not contribute to 
the commutator, we get as our final result 
$$\langle [x_u,\sigma_t x_t] \rangle_{AV} 
=i_{\rm eff} \hbar \omega_{ut} \sigma_t-\tau\langle {\dot x}_u^{\prime}
{\rm Tr} \tilde C \sigma_t x_t^{\prime} \rangle_{AV}~~~.\eqno(28a)$$
Equation (28a), like Eq.~(20b) of the preceding section that it generalizes, 
is exact.  

In order to have emergent quantum behavior, it is necessary that the 
second term on the right hand side of Eq.~(28a) be much smaller than the 
first term, that is, we require 
$$\hbar^{-1} \tau |\langle {\dot x}_u^{\prime}
{\rm Tr} \tilde C \sigma_t x_t^{\prime} \rangle_{AV}| <<1~~~. \eqno(28b)$$
Again, by replacing $\tilde C$ by its expectation value at the 
start of the derivation leading to Eq.~(28a), we get a second Ward 
identity 
$$0
=i_{\rm eff} \hbar \omega_{ut} \sigma_t-\tau\langle {\dot x}_u^{\prime}
{\rm Tr}  i_{\rm eff} \hbar \sigma_t x_t^{\prime} \rangle_{AV}~~~
.\eqno(28c)$$
Hence the inequality of Eq.~(28b) can be satisfied only if $\tilde C$ is 
not an extensive quantity, since if $\tilde C$ were extensive one could, 
in the large system limit, approximate  it in Eq.~(28b) by its expectation 
$i_{\rm eff} \hbar$, giving an expression that, by Eq.~(28c), cannot 
be small.  

Because the derivation of this section is valid for fermions 
as well as bosons, one can in 
fact make a stronger statement about the conditions for emergent quantum 
behavior.  Letting the indices $t$ and $u$ in Eq.~(28a) be either both 
bosonic or both fermionic, we get the respective relations 
$$\eqalign{
\langle [q_r,p_r]\rangle_{AV}=&i_{\rm eff}\hbar 
-\tau\langle{\dot q}_r^{\prime}
{\rm Tr}\tilde C p_r^{\prime}\rangle_{AV}~~~~~r~{\rm bosonic}\cr
\langle\{q_r,p_r\}\rangle_{AV}=&i_{\rm eff}\hbar
-\tau\langle{\dot q}_r^{\prime}
{\rm Tr}\tilde C p_r^{\prime}\rangle_{AV}~~~~~r~{\rm fermonic}~~~.\cr
}\eqno(29a)$$
Substituting this into Eq.~(7) for $\tilde C$, taking the ensemble average, 
and using Eq.~(16c), we get 
$$\eqalign{
i_{\rm eff}\hbar =& \langle \tilde C \rangle_{AV}
=\langle  \sum_{r~{\rm bosons}}[q_r,p_r]-\sum_{r~{\rm fermions}}
\{q_r, p_r\} \rangle_{AV}\cr
=&\left(\sum_{r~{\rm bosons}}-\sum_{r~{\rm fermions}} \right) 
i_{\rm eff}\hbar
-\tau \left(\sum_{r~{\rm bosons}}-\sum_{r~{\rm fermions}} \right) 
\langle{\dot q}_r^{\prime}{\rm Tr}\tilde C p_r^{\prime}\rangle_{AV} ~~~,\cr
}\eqno(29b)$$
which on division by $\hbar$ and transposition of terms gives 
$$ \left(\sum_{r~{\rm bosons}}-\sum_{r~{\rm fermions}} \right) 
\hbar^{-1} \tau\langle{\dot q}_r^{\prime}{\rm Tr}\tilde C 
p_r^{\prime}\rangle_{AV} \break
=i_{\rm eff} \left(\sum_{r~{\rm bosons}}-\sum_{r~{\rm fermions}}-
1 \right)~~~.
\eqno(29c)$$
When the condition of Eq.~(28b) for emergent canonical behavior is 
satisfied, the left hand side of Eq.~(29c) is a sum of very small terms. 
Assuming that this sum yields at most a finite, bounded total, let us  
consider the case in which $r$ includes the spatial label 
of a translation invariant field theory.  Then 
the number of bosonic and fermionic modes per unit volume contributing on 
the right hand side of Eq.~(29c) must be equal, since if not, 
the right hand side of Eq.~(29c) 
would become infinite as the spatial volume grows to infinity, contradicting 
the boundedness of the left hand side.  Therefore, in a complex Hilbert 
space\footnote{*}{For a discussion of how our arguments must be modified 
in real and quaternionic Hilbert space, see the Appendix.}, a candidate 
pre-quantum mechanics theory must have equal numbers 
of bosonic and fermionic degrees of freedom, making it plausible that 
such a candidate theory should be supersymmetric.  When the numbers of 
bosonic and fermionic modes are in balance, Eq.~(29c) simplifies to 
$$ \left(\sum_{r~{\rm bosons}}-\sum_{r~{\rm fermions}} \right) 
\hbar^{-1} \tau\langle{\dot q}_r^{\prime}{\rm Tr}\tilde C 
p_r^{\prime}\rangle_{AV} \break
=-i_{\rm eff}~~~, \eqno(29d)$$
showing that the remainder terms in Eq.~(28a), that are neglected when 
Eq.~(28b) is satisfied, sum in Eq.~(29c) to give a total of unit magnitude.  
                
Corresponding to the Ward identity of Eq.~(28a), we can derive a class 
of more general Ward identities by replacing $\sigma_t x_t^{\prime}$ 
in Eq.~(25b) by a general $U$, constructed as a Weyl ordered  
(i.e., symmetrized) polynomial in the 
products $\{ \sigma_r x_r\}$, with coefficients that are $c$-number 
functions of 1 and of $i_{\rm eff}$.  
In place of Eq.~(27b), we now get 
$$\eqalign{
0=&\langle (-\tau \sum_r {\dot x}_r^{\prime} \omega_{rs} 
-[\tilde \lambda,\sum_r \omega_{rs} x_r^{\prime}]) 
{\rm Tr} \tilde C U \cr
+&[U,\sum_r \omega_{rs} x_r^{\prime}]  
+  \sum_{\rm each~{\it x_s}~in~{\it U}}  
U({\rm one}~x_s \to \tilde C)^{\prime}\rangle_{AV} ~~~.\cr
}\eqno(30a)$$
As long as $U$ has coefficients that depend only on the matrices (or 
operators) 
1 and $i_{\rm eff}$, the second 
term on the right in Eq.~(30a), which involves 
a commutator with $\tilde \lambda$, 
vanishes by the same arguments as before.  Contracting the remainder with 
$\omega_{us}$, but making no approximations, we get as the exact general Ward 
identity analogous to Eq.~(28a), 
$$\langle [x_u,U] \rangle_{AV} 
=\langle \sum_s \omega_{us} 
\sum_{\rm each~{\it x_s}~in~{\it U}}  
U({\rm one}~x_s \to \tilde C)^{\prime}\rangle_{AV} 
-\tau\langle {\dot x}_u^{\prime}
{\rm Tr} \tilde C U \rangle_{AV}~~~.\eqno(30b)$$
Suppose now that we can make the following two approximations,  
(i) we replace $\tilde C$ in the 
first term on the right hand side of Eq.~(30b) by its ensemble average 
$i_{\rm eff} \hbar$, and   
(ii) we 
neglect the $\tau$ term in Eq.~(30b). 
We then are left with the relation 
$$\langle [x_u,U] \rangle_{AV} 
=\langle \sum_s \omega_{us} 
 \sum_{\rm each~{\it x_s}~in~{\it U}}  
U({\rm one}~x_s \to i_{\rm eff} \hbar^{\prime})\rangle_{AV} 
~~~,\eqno(30c)$$
which extends the effective canonical algebra inside ensemble averages 
to the commutator                        
of $x_u$ with a general Weyl ordered polynomial $U$.  

By the methods of 
Appendix E of [4], Equation~(30c) can be 
extended to include sources for the remainder parts $x_r^{\prime}$ of the  
phase space variables.  Specializing  
the relations obtained this way to the case $U=i_{\rm eff}$ implies 
that $i_{\rm eff}$ can be freely commuted with phase space variables 
inside ensemble averages. As argued in [4],    
the resulting set of Ward identities then yields the
canonical generator structure, including the time evolution 
relations, of Heisenberg picture quantum field theory.  The only assertion 
in [4] that cannot be derived this way is the claim that the 
time evolution equation is more exact than the other generator relations; 
this claim used the assumption that $\tilde C$ can be replaced by its 
ensemble average inside the $\tau$ term, which we have seen is not correct.  
The remainder of the 
conclusions of [4] rest on the two approximations that we made above, 
which can be rephrased as the assumptions that, (i) in the terms of 
Eq.~(30b) that involve the {\it unvaried} canonical ensemble $\rho$ of 
Eq.~(13) with a factor of $\tilde C$ in the integrand, 
the fact that the ensemble is sharply peaked around the mean allows us 
to replace the integrand factor $\tilde C$ by its ensemble 
average $i_{\rm eff} \hbar$, and (ii)  the canonical ensemble displays a 
certain rigidity, in  that terms of the form 
$\int d\mu \delta \rho {\rm Tr} \tilde C U$  can be dropped.  On the other 
hand, we have seen that terms of the form $\int d\mu \delta \rho {\rm Tr} 
i_{\rm eff} \hbar U$ cannot be dropped; this does not contradict our 
assumptions 
because $\delta \rho$ can be rapidly varying around the peak of the 
ensemble.  

\bigskip                                                  
\centerline{\bf 5.  ~~Discussion}

We have seen that a statistical mechanics can be formulated for a wide 
class of matrix models with a global unitary invariance, and that within 
this statistical mechanics, the ensemble averages of canonical variables
obey the exact relations of Eq.~(20b), (28a), and (29a), and (30b).  
When these can 
be approximated by dropping the $\tau$ terms [and, in Eq.~(30b), replacing 
the $\tilde C$ insertions in $U^{\prime}$ by their ensemble averages],  
the result is emergent 
quantum mechanical behavior for the statistical ensemble averages. 
The condition for validity of the approximation of neglecting the $\tau$ 
terms is rather delicate:  we have argued that it requires that 
$\tilde C$  should be an intensive thermodynamic quantity, and that 
the numbers of bosonic and fermionic degrees of freedom should balance.  

We conclude with some brief remarks:

\item{[1]}  In the first two references of [3], it is shown that one can 
readily formulate trace dynamics models in which global unitary operator 
invariance 
is gauged to give a local unitary operator invariance.  Since global unitary 
invariance is a special case of local unitary invariance, the considerations 
of this paper apply to these models.  
 
\item{[2]}  In Refs. [5], it is shown that supersymmetric Yang Mills 
theory, and the related ``matrix model for $M$ theory'', fit 
naturally into the 
trace dynamics framework analyzed in this paper.  In these models, $\tilde C$ 
vanishes up to a surface term contribution, a behavior consistent with its 
being an intensive thermodynamic quantity.  

\item{[3]}  Although our final results of Eqs.~(20b), (28a), and (29a) 
superficially resemble the string-inspired formula of Eq.~(1b), there is 
an important difference.  In Eq.~(1b) the coordinate is a quantum operator, 
as is usual in nonrelativistic quantum mechanics, 
whereas in our results of Eqs.~(20b), etc., the coordinate is merely a 
degree of freedom label $r$, as it always is in quantum field theory, and 
the coordinates and momenta are canonical field variables with label $r$.  
It may be possible to make a connection between the two types of modified 
commutation relations when the metric structure of the coordinate manifold 
is taken into account, using the fact that the proper distance is related 
to the coordinate interval by $ds^2=g_{\mu\nu}dx^{\mu}dx^{\nu}$.  In a field 
theoretic interpretation, $dx^{\mu}$ is just the change in the degree of 
freedom label, but the metric  is a dynamical variable, and hence so is 
$ds$.  This suggests that there may be an analog of Eq.~(1b) involving the 
relativistic proper distance, and that this is the relation to be compared 
with our results in this paper.  

\item{[4]}  As is well known, in a complex Hilbert space the canonical 
algebra $[q,p]=i,~ [q,i]=[p,i]=0$ cannot have finite dimensional 
representations, since this algebra implies, for example,  the relation
$q^2p^2+p^2q^2-2qp^2q=-2$, which in a finite dimensional Hilbert space  
would have a left hand side with trace zero and a right hand side with 
trace nonzero.  However, it is consistent for the canonical algebra 
to emerge as the limit $N \to \infty$ of an algebra in an $N$ dimensional  
Hilbert space, 
which is the behavior argued for in Ref. [4] and here.  Because the 
emergent canonical algebra involves not the imaginary unit $i$ of the 
underlying complex Hilbert space, but rather the operator $i_{\rm eff}$ 
with ${\rm Tr} i_{\rm eff}=0$, a basis for the operator algebra in 
the emergent theory is provided by 
a set of operators that commute with $i_{\rm eff}$, together 
with one additional operator that anticommutes with $i_{\rm eff}$, and that 
plays the role of the time reversal operator in the emergent complex quantum 
mechanics.  In fact, because the condition ${\rm Tr} i_{\rm eff}=0$  implies 
that one can find a representation in which $i_{\rm eff}$ is a real matrix 
(just as for Pauli matrices $\rho_{1,2,3}$ the matrix  $i\rho_3$ can be 
given the real form $i \rho_2$ 
by a change of representation), the quantum mechanics emergent from 
matrix dynamics has the structure of a complexified real 
quantum  mechanics, for which the operator algebra  
has the form just described (see, e.g., Sec. 2.6 of the 
second citation in Ref. [3]).  

\item{[5]}  We have seen that the emergence of quantum mechanical behavior 
from matrix model statistical mechanics requires a certain ``rigidity'' of 
the 
statistical ensemble.  It is easy to see [8, 9] 
that this rigidity is a sufficient 
condition for the canonical and microcanonical ensembles to give the same 
Ward identities, and hence the same emergent quantum behavior.  The 
need for a rigid statistical ensemble in our context suggests a possible 
analogy with the concept of London rigidity in the theory of 
superconductivity [10].  In the presence of an applied vector potential 
$\vec A$, the 
induced current density in a metal is given by 
$$\langle \vec j ~\rangle = -{n e \over m} \langle \vec p + {e \over c} 
\vec A
~\rangle ~~~.\eqno(31)$$
In a normal metal the two terms on the right hand side of Eq.~(31a) nearly 
cancel, leaving a small residual diamagnetism.  However, in a superconductor 
the rigidity of the wave 
function leads to the vanishing of $\langle \vec p~ \rangle$, giving perfect 
diamagnetism and the Meissner effect.   An analogy with the  
results of this paper would equate normal metal behavior with the case in 
which $\tilde C$ can be replaced by its ensemble average in the $\tau$ term;  
in this case the right hand side of Eq.~(20b) is approximately equal to the 
right hand side of Eq.~(22b), leading to cancellation of the emergent 
canonical commutator.  Similarly,  
the analogy would equate superconducting behavior 
with the case in which the $\tau$ term containing $\tilde C$ can be dropped 
because of ``rigidity'' of $\delta \rho$, leading through Eqs.~(20b), 
(28a), and (29a) to an emergent canonical commutator as an analog of the 
superconductive Meissner effect.  
\bigskip                                                           
\centerline{\bf Acknowledgments}
The work of S.L.A. was supported in part by the Department of Energy under
Grant \#DE--FG02--90ER40542, and he wishes to acknowledge a stimulating   
conversation with Walter Troost.  He also wishes to acknowledge 
the hospitality of the   
Aspen Center for Physics, where this work was completed, and of the  
Department of Applied Mathematics and Theoretical Physics at Cambridge 
University.  A. K. in turn wishes to acknowledge the hospitality of the 
Institute for Advanced Study in Princeton.  
\bigskip
\centerline{\bf Appendix: Real and Quaternionic Hilbert Space}

For reasons that we now describe, in real and quaternionic Hilbert space 
the arguments of this paper must be modified and yield  weaker 
conclusions.  The underlying reason for this modification is that only 
in complex Hilbert space can one have a non-real trace that nonetheless 
obeys the cyclic property.   In quaternionic Hilbert space,  as a consequence 
of the noncommutativity of the quaternions, only the real part of the 
ordinary trace obeys the cyclic property.  In real Hilbert space, the 
trace is necessarily real, and the trace of any anti-self-adjoint 
operator vanishes.   Thus in these two cases, if one follows [3, 4] 
and defines the 
graded trace ${\bf Tr}$ to be the one that obeys the cyclic property, then  
the definition must include taking the real part, and to get a nonzero 
result one must require the operator argument $V$ of ${\bf Tr}V$ 
to be self-adjoint.  As a consequence, in the general case 
derivation analogous to that starting from Eq.~(25b),  one must consider  
${\bf Tr}\{i_{\rm eff}, \tilde C\} \sigma_t x_t^{\prime}$ rather than 
${\rm Tr} \tilde C \sigma_t x_t^{\prime}$. 
This gives the following analog of Eq.~(28a), 
$$\langle [x_u,\sigma_t \{i_{\rm eff},x_t\}] \rangle_{AV} 
=\{i_{\rm eff},i_{\rm eff} \hbar\}^{\prime} \omega_{ut} \sigma_t
-\tau\langle {\dot x}_u^{\prime}
{\rm Tr} \{i_{\rm eff},\tilde C\} 
\sigma_t x_t^{\prime} \rangle_{AV}~~~,\eqno(A1)$$
where the prime on the first term on the right hand side of Eq.~(A1) 
indicates extraction of the traceless part.  However, since the traceless
part of $\{i_{\rm eff}, i_{\rm eff} \hbar \}= -2 \hbar$ is zero, Eq.~(A1) 
becomes 
$$\langle [x_u,\sigma_t \{i_{\rm eff},x_t\}] \rangle_{AV} 
=-\tau\langle {\dot x}_u^{\prime}
{\rm Tr} \{i_{\rm eff},\tilde C\} 
\sigma_t x_t^{\prime} \rangle_{AV}~~~,\eqno(A2)$$
which has a structure more like the equation obtained by 
subtracting Eq.~(28c) from Eq.~(28a) than like Eq.~(28a) itself.      
In complex  
Hilbert space, Eqs.~(A2) {\it and} Eqs.~(28a, c) all hold, and conditions 
for emergent quantum behavior can be formulated from the latter as in the 
text; in the 
real and quaternionic Hilbert space cases, only Eq.~(A2) holds, and we 
cannot proceed with the analysis of the text.  

The remarks made here 
correct a subtle error made in Sec.~6 of Adler and Millard [4].  There, 
the classical sources introduced in this paper were set to zero, 
and Ward identities 
were derived by varying $x_s$, not just the noncommutative part 
$x_s^{\prime}$ 
as we have done here.  This gives as the Ward identity analogous to  
Eq.~(A1) 
$$\langle [x_u,\sigma_t \{i_{\rm eff},x_t\}] \rangle_{AV} 
=\{i_{\rm eff},i_{\rm eff} \hbar\} \omega_{ut} \sigma_t
-\tau\langle {\dot x}_u
{\rm Tr} \{i_{\rm eff},\tilde C\} 
\sigma_t x_t \rangle_{AV}~~~.\eqno(A3)$$  
Equation (A3) is a valid relation, but it actually implies {\it two} 
relations, quite different in structure, for its classical or $c$-number part 
and its primed or traceless part.  Separating Eq.~(A3) into these parts, and 
ignoring interference terms between $x^c$ and $x^{\prime}$ in the $\tau$ 
term, one finds that the 
primed part of Eq.~(A3) gives Eq.~(A2), while the 
$c$-number part of Eq.~(A3) becomes an equipartition identity for $x^c$ and 
gives no direct information about the expectation of the commutator.  
Hence the arguments of [4] do not, in their present form, 
provide evidence for emergent                                    
quantum behavior in the real and quaternionic Hilbert space cases.  

\vfill\eject
\centerline{\bf References}
\bigskip
\noindent
\item{[1]}  See, e.g., D. J. Gross and P. F. Mende, Nucl. Phys. 
B303 (1988) 407 ; D. Amati, M. Cialfaloni, and G. Veneziano, Phys. Lett. 
B216 (1989) 41; M. Maggiore, Phys. Lett. B319 (1993) 83 ; recent reviews 
are E. Witten, Phys. Today 49(4) (1996) 24 ; L. J. Garay, Int. J. 
Mod. Phys. A10 (1995) 145.
\bigskip
\noindent
\item{[2]} A. Kempf, J. Math. Phys. 35 (1994) 4483 ; A. Kempf, 
G. Mangano, and R. B. Mann, Phys. Rev. D52 (1995) 1108 ; 
A. Kempf and H. Hinrichsen, J. Math. Phys. 37 (1996) 2121; 
A. Kempf, preprint DAMTP/94-33, hep-th/9405067;  A. Kempf, J. Math Phys. 
38 (1997) 1347; A. Kempf, Phys. 
Rev. D54 (1996) 5174; A. Kempf and G. Mangano, Phys. Rev. D55 (1997) 7909; 
A. Kempf, preprint DAMTP/96-101, hep-th/9612082, to appear 
in Proc. XXI Int. Coll. on Group Theor. Methods in Physics, Goslar, 
July 1996.
\bigskip
\noindent
\item{[3]}  S. L. Adler, Nucl. Phys. B 415 (1994) 195; S. L. Adler, 
``Quaternionic Quantum Mechanics and Quantum Fields,'' Secs. 13.5-13.7 and 
App. A (Oxford Univ. Press, New York, 1995);
S. L. Adler, G. V. Bhanot, and J. D. Weckel, J. Math. Phys. 
35 (1994) 531; S. L. Adler and Y.-S. Wu, Phys. Rev. D49 (1994) 6705.  
\bigskip
\noindent
\item{[4]}  S. L. Adler and A. C. Millard, Nucl. Phys. B 473 (1996) 199. 
\bigskip
\noindent
\bigskip
\noindent
\item{[5]} S. L. Adler, preprint IASSNS-HEP-97/15, hep-th/9703053, Phys. 
Lett. 
B {in press}; 
S. L. Adler, preprint IASSNS-HEP-97/16, hep-th/9703132, 
Nucl. Phys. B (in press).  
\bigskip
\noindent
\item{[6]}  B. de Wit, J. Hoppe, and H. Nicolai, Nucl. Phys. B305 (1988) 545;  
P. K. Townsend, Phys. Lett. B373 (1996) 68; 
T. Banks, W. Fischler, S. H. Shenker, and L. Susskind, ``M Theory as A 
Matrix Model: A Conjecture'', 
hep-th/9610043.  This paper gives an extensive bibliography, and has 
inspired a large number of very recent 
preprints on the relationship between string theory and matrix models.  
For further bibliography, see the first paper of Ref.~[5] above.  
\bigskip
\noindent
\item{[7]}  A. C. Millard, ``Non-Commutative Methods in Quantum 
Mechanics'', Princeton University PhD thesis, April, 1997. 
\bigskip
\noindent
\item{[8]}  S. L. Adler and   L. P. Horwitz, J. Math. Phys. 37 (1996) 5429.
\bigskip
\noindent
\item{[9]}  S. L. Adler and L. P. Horwitz, unpublished.
\bigskip
\noindent
\item{[10]}  J. R. Schrieffer, ``Theory of Superconductivity'', Sec. 8-1,  
p. 204 (W. A. Benjamin, New York, 1964).  
\bigskip
\noindent
\bigskip
\noindent
\bigskip
\noindent
\bigskip
\noindent
\bigskip
\noindent
\bigskip
\noindent
\bigskip
\noindent
\bigskip
\noindent
\bigskip
\noindent
\bigskip
\noindent
\bigskip
\noindent
\bigskip
\noindent
\vfill
\eject
\bigskip
\bye